\pdfoutput=1
\documentclass[preprint,preprintnumbers,amsmath,amssymb,floatfix,endfloats*]{revtex4}

\usepackage{graphicx}
\usepackage{dcolumn}
\usepackage{bm}
\usepackage{amsfonts}

\DeclareMathAlphabet{\mathsfsl}{OT1}{cmr}{bx}{it}
\begin{document}
\title{Slow relaxation dynamics in binary glasses during stress-controlled, tension-compression cyclic loading}
\author{Nikolai V. Priezjev$^{1,2}$}
\affiliation{$^{1}$Department of Mechanical and Materials
Engineering, Wright State University, Dayton, OH 45435}
\affiliation{$^{2}$National Research University Higher School of
Economics, Moscow 101000, Russia}
\date{\today}
\begin{abstract}

The effect of cyclic loading on relaxation dynamics and mechanical
properties of metallic glasses is studied using molecular dynamics
simulations. We consider the Kob-Andersen three-dimensional binary
mixture rapidly cooled across the glass transition and subjected to
thousands of tension-compression cycles in the elastic range.  It
was found that during cyclic loading at constant pressure, the
system is relocated to progressively lower levels of the potential
energy, thus promoting greater densification and higher strength.
Furthermore, with increasing stress amplitude, the average glass
density increases and the minimum of the potential energy becomes
deeper, while the elastic modulus is reduced. The typical size of
clusters of atoms with large nonaffine displacements becomes smaller
over consecutive cycles, which correlates with the gradual decrease
in the potential energy. These results are important for
thermomechanical processing of metallic glasses with improved
mechanical properties.

\vskip 0.5in

Keywords: glasses, deformation, temperature, strain amplitude,
molecular dynamics simulations

\end{abstract}

\maketitle

\section{Introduction}

Understanding the quantitative correlation between atomic structure
and mechanical and physical properties of amorphous materials such
as metallic glasses is important for various structural and
biomedical applications~\cite{WangRev12}. It is commonly accepted
that deformation of amorphous materials occurs via a series of
sudden rearrangements of small groups of particles often called
shear transformation zones~\cite{Argon79,Spaepen77,Falk98}.
Interestingly, it was recently demonstrated that the shear modulus
as well as local shear transformations can be accurately predicted
using the `flexibility volume', a parameter that combines the
information of both static atomic volume and thermal
vibrations~\cite{EvanMa16}. The results of atomistic simulations of
cyclic nanoindentation of metallic glasses have revealed hardening
effects that arise due to local stiffening and densification of the
region containing the original yielding
path~\cite{Schuh08,Schuh12,WangASS17,Lashgari18}. However, the
influence of complex deformation protocol and sample preparation
history on fatigue lifetime, structural changes, and yielding
requires further investigation.

\vskip 0.05in

Recent molecular dynamics simulation studies of the fatigue
mechanism in metallic glasses under tension-compression,
strain-controlled cyclic loading revealed that a shear band is
initiated at the sample surface due to a slow accumulation of shear
transformation zones, followed by a relatively quick propagation of
a shear band across the sample~\cite{GaoNano15,GaoSize17}.   In
contrast, the results of atomistic simulations of metallic glasses
under the uniaxial stress-control mode have shown that the plastic
deformation in small-size samples proceeds via network-like
shear-transition zones~\cite{Liaw10}. Furthermore, it was
demonstrated that low-cycle fatigue tests on metallic glass
nanowires result in work hardening or softening, depending on the
applied load, and, after strain-dependent microscopic damage
accumulation, a shear band forms rapidly~\cite{Shi11,Shi15}.  More
recently, it was also found that well-annealed metallic glasses,
subjected to periodic compressive stress below yield, exhibit
softening with either decreasing cycling frequency or increasing
stress amplitude~\cite{Yang16}.  Despite extensive efforts, however,
how exactly the yielding transition and shear band formation depend
on the cyclic loading conditions (stress versus strain controlled),
frequency, and system size remains not fully understood.

\vskip 0.05in

In recent years, the mechanical response of amorphous materials
subjected to periodic shear was extensively studied using either
athermal,
quasistatic~\cite{Sastry13,Reichhardt13,IdoNature15,Kawasaki16,Sastry17,OHern17,Lavrentovich17}
or finite
temperature~\cite{Priezjev13,Priezjev14,Priezjev16,Priezjev16a,Priezjev17,Hecke17,Keblinsk17,Priezjev18,Priezjev18a}
molecular dynamics simulations.  Under cyclic loading, athermal
systems were shown to settle into the so-called limit cycles, where
all particles return to their positions after one or several shear
cycles~\cite{Reichhardt13,IdoNature15}. Moreover, the number of
shear cycles required to reach a steady state increases upon
approaching the yield strain from
below~\cite{Reichhardt13,Sastry13,IdoNature15}. Above the yield
point, periodic deformation of well-annealed glasses results in
sudden formation of a system-spanning shear band after a number of
transient cycles~\cite{GaoNano15,Priezjev17,Sastry17}. Remarkably,
it was recently demonstrated that in contrast to dense amorphous
systems, where the yielding transition is associated with a drop of
the elastic modulus and enhanced particle diffusion, weakly jammed
solids exhibit two-step yielding process; namely, softening at
intermediate strain amplitudes and the onset of diffusion at
distinctly larger yield strain~\cite{Hecke17}.

\vskip 0.05in

In this paper, we investigate mechanical properties and structural
relaxation dynamics of poorly annealed binary glasses under periodic
tension-compression, stress-controlled loading using molecular
dynamics simulations.  It will be shown that during several thousand
cycles in the elastic range, the amorphous systems continue to
explore progressively lower potential energy states and become more
dense. The slow relaxation process is associated with small-scale
irreversible particle rearrangements that are facilitated by the
applied loading. As the stress amplitude increases, the average
glass density becomes larger and both the potential energy and
elastic modulus are reduced.

\vskip 0.05in

The rest of the paper is structured as follows.  The details of the
molecular dynamics simulation model and the stress-controlled cyclic
loading protocol are described in the next section. The numerical
results for the time dependence of the potential energy, glass
density, and elastic modulus as well as the analysis of nonaffine
displacements are presented in Sec.\,\ref{sec:Results}. The main
results are briefly summarized in the last section.

\section{Details of MD simulations}
\label{sec:MD_Model}

Molecular dynamics simulations are carried out on a glass former
that consists of the non-additive binary mixture (80:20) first
introduced by Kob and Andersen (KA)~\cite{KobAnd95} to study
properties of the amorphous metal alloy
$\text{Ni}_{80}\text{P}_{20}$~\cite{Weber85}. In this model, the
interaction between atoms of types $\alpha,\beta=A,B$ is given by
the pairwise Lennard-Jones (LJ) potential:
\begin{equation}
V_{\alpha\beta}(r)=4\,\varepsilon_{\alpha\beta}\,\Big[\Big(\frac{\sigma_{\alpha\beta}}{r}\Big)^{12}\!-
\Big(\frac{\sigma_{\alpha\beta}}{r}\Big)^{6}\,\Big],
\label{Eq:LJ_KA}
\end{equation}
where $\varepsilon_{\alpha\beta}$ and $\sigma_{\alpha\beta}$ are the
energy and length scales of the LJ potential~\cite{Allen87}.   The
interaction parameters for both types of atoms are chosen as
follows: $\varepsilon_{AA}=1.0$, $\varepsilon_{AB}=1.5$,
$\varepsilon_{BB}=0.5$, $\sigma_{AB}=0.8$, $\sigma_{BB}=0.88$, and
$m_{A}=m_{B}$~\cite{KobAnd95}.  In addition, the cutoff radius is
set to $r_{c,\,\alpha\beta}=2.5\,\sigma_{\alpha\beta}$ to improve
computational efficiency. All physical quantities are expressed in
the reduced LJ units of length, mass, energy, and time, which are
defined as $\sigma=\sigma_{AA}$, $m=m_{A}$,
$\varepsilon=\varepsilon_{AA}$, and
$\tau=\sigma\sqrt{m/\varepsilon}$, respectively.  The total number
of atoms throughout the study is $N_{tot}=40\,000$. The equations of
motion were integrated using the velocity Verlet
algorithm~\cite{Allen87} with the time step $\triangle
t_{MD}=0.005\,\tau$ using the LAMMPS numerical code~\cite{Lammps}.

\vskip 0.05in


We next describe the preparation procedure and the stress-controlled
deformation protocol at constant pressure.  The system was initially
equilibrated at the temperature $1.1\,\varepsilon/k_B$ and constant
volume, which corresponds to the atomic density
$\rho=\rho_{A}+\rho_{B}=1.2\,\sigma^{-3}$. Here, $k_B$ denotes the
Boltzmann constant.  This temperature is well above the glass
transition temperature $T_c\approx0.435\,\varepsilon/k_B$ of the KA
model at the density $\rho=1.2\,\sigma^{-3}$~\cite{KobAnd95}. In the
next step, the temperature was instantaneously reduced to
$T_{LJ}=0.1\,\varepsilon/k_B$ below the glass transition, and the
pressure along the $\hat{x}$ and $\hat{y}$ directions was set to
zero.  In contrast, the normal stress along the $\hat{z}$ direction
was varied periodically with the period of oscillations
$T=500\,\tau$.   More specifically, the normal stress $\sigma_{zz}$
was changed piecewise linearly as follows: from 0 to
$\sigma_{zz}^{max}$ during $T/4$, from $\sigma_{zz}^{max}$ to
$-\sigma_{zz}^{max}$ during $T/2$, and from $-\sigma_{zz}^{max}$
back to $0$ during $T/4$.   Thus, the average of $\sigma_{zz}$ over
the oscillation period is zero, and the system was subjected to
$5000$ stress-controlled, tension-compression cycles. In all
simulations, periodic boundary conditions were applied along the
$\hat{x}$, $\hat{y}$, and $\hat{z}$ directions. The data for the
potential energy, stresses, system dimensions, and atomic
configurations were collected in one sample for the postprocessing
analysis.

\section{Results}
\label{sec:Results}



\begin{figure}[t]
\includegraphics[width=12.cm,angle=0]{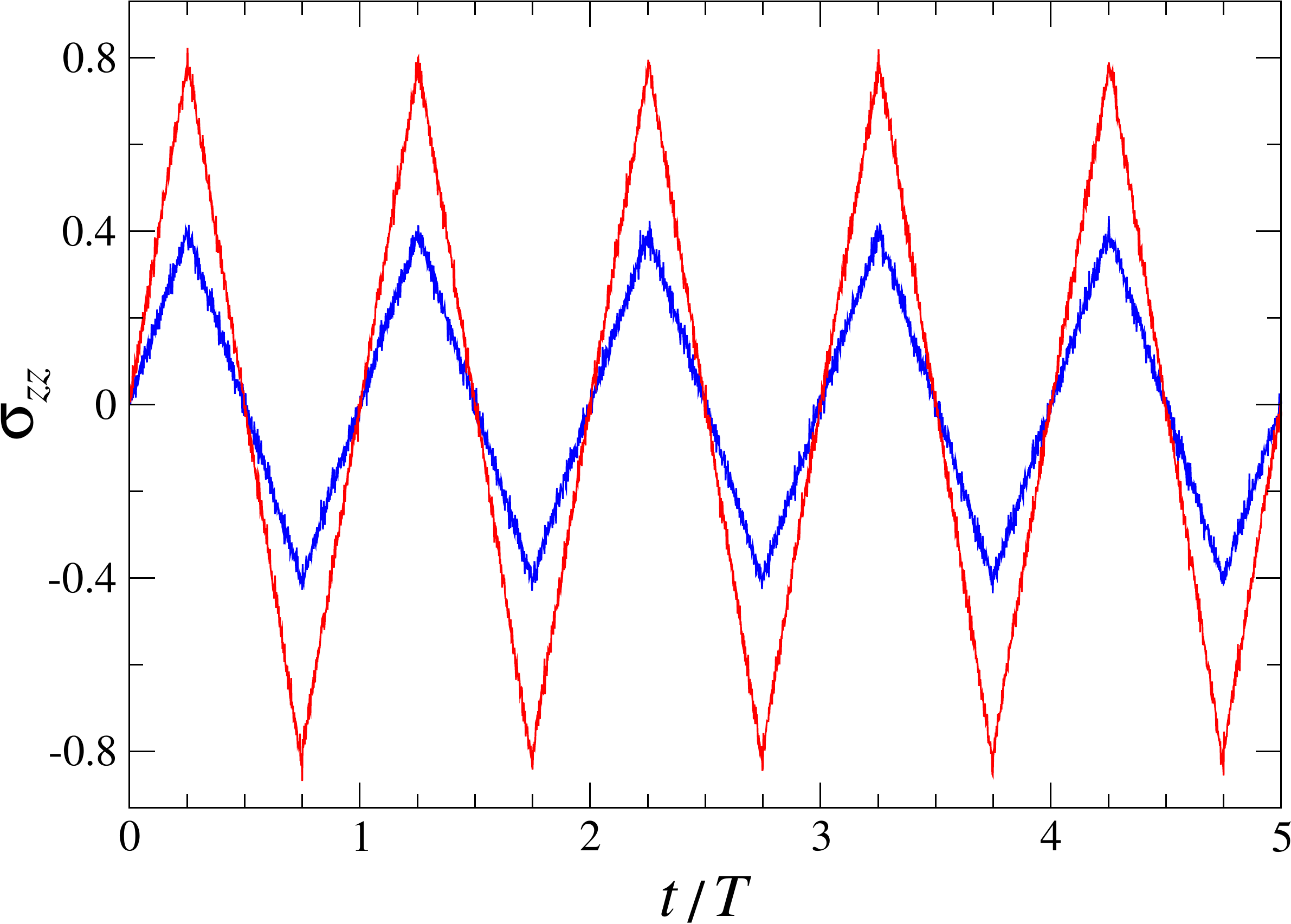}
\caption{(Color online) The time series of the normal stress
$\sigma_{zz}$ (in units of $\varepsilon\sigma^{-3}$) during the
first 5 periods after the thermal quench to the temperature
$T_{LJ}=0.1\,\varepsilon/k_B$.  The stress amplitude is
$\sigma_{zz}^{max}=0.4\,\varepsilon\sigma^{-3}$ (blue lines) and
$\sigma_{zz}^{max}=0.8\,\varepsilon\sigma^{-3}$ (red lines). The
oscillation period is $T=500\,\tau$. }
\label{fig:stress_zz_5T_Tr1}
\end{figure}

As described in the previous section, the normal stress in the
$\hat{z}$ direction was applied piecewise linearly in time with the
oscillation period $T=500\,\tau$.  As an example, the actual stress
$\sigma_{zz}$ measured in MD simulations during the first $5$ cycles
after the thermal quench to the temperature
$T_{LJ}=0.1\,\varepsilon/k_B$ is presented in
Fig.\,\ref{fig:stress_zz_5T_Tr1} for two values of the stress
amplitude.   As expected, it can be seen that the stress variation
is linear with superimposed fluctuations during the consecutive time
intervals. Note also that in the case
$\sigma_{zz}^{max}=0.8\,\varepsilon\sigma^{-3}$, the stress becomes
slightly smaller than the value $-0.8\,\varepsilon\sigma^{-3}$, when
the sample is at maximum tension.   It should be emphasized that the
current MD setup with periodic boundary conditions imposed along the
transverse directions to loading is not designed to study extended
plastic deformation like shear bands.   In order to observe the
formation of shear bands, which typically run at a certain angle
with respect to the loading direction, the simulation cell must have
non-periodic boundary conditions in at least one of the transverse
directions~\cite{GaoNano15}.

\vskip 0.05in

\begin{figure}[t]
\includegraphics[width=12.cm,angle=0]{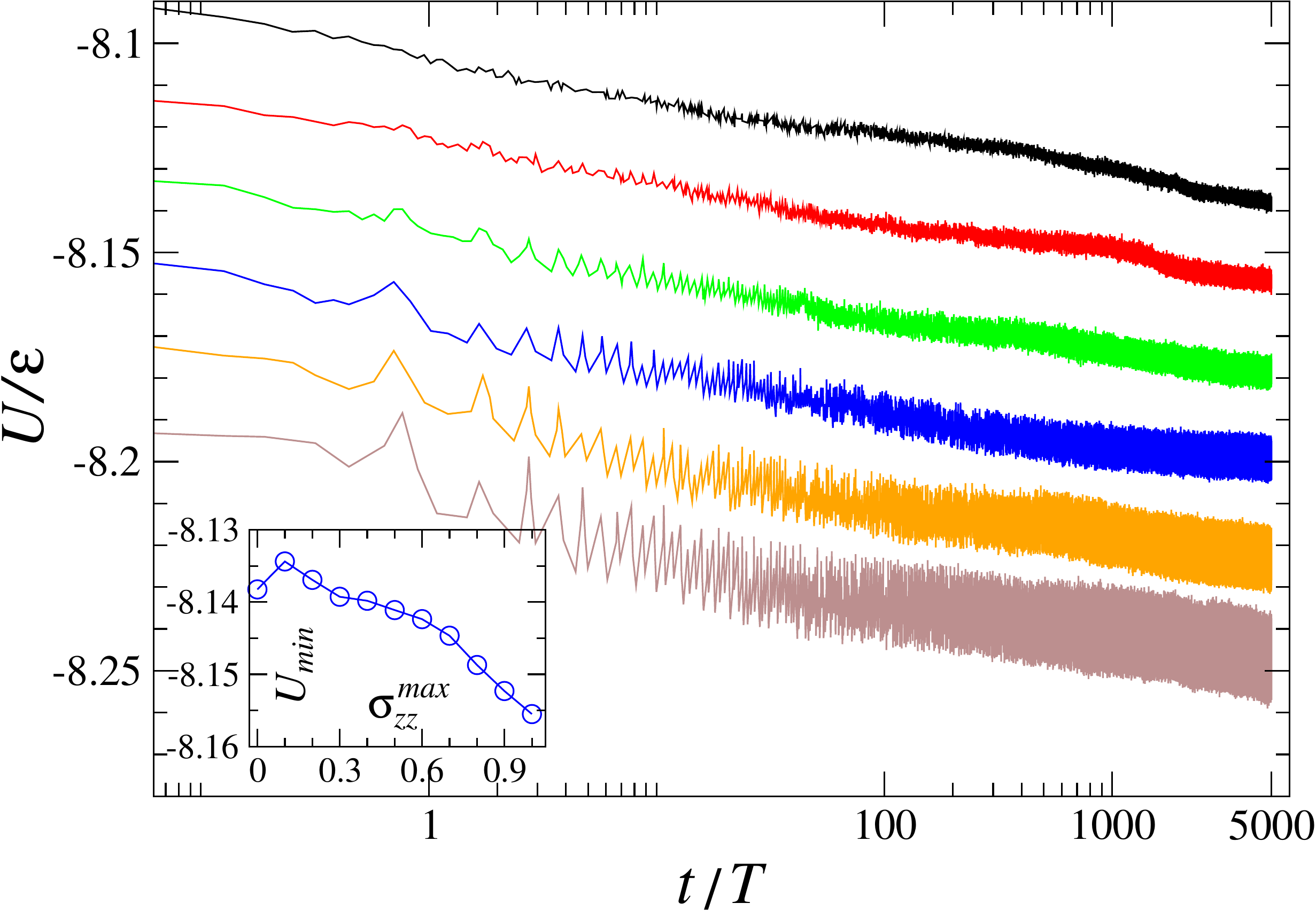}
\caption{(Color online) The potential energy per atom
$U/\varepsilon$ during 5000 tension-compression cycles after the
thermal quench for different stress amplitudes.  The data are
displaced by $-0.02\,\varepsilon$ for
$\sigma_{zz}^{max}=0.2\,\varepsilon\sigma^{-3}$ (red), by
$-0.04\,\varepsilon$ for
$\sigma_{zz}^{max}=0.4\,\varepsilon\sigma^{-3}$ (green), by
$-0.06\,\varepsilon$ for
$\sigma_{zz}^{max}=0.6\,\varepsilon\sigma^{-3}$ (blue), by
$-0.08\,\varepsilon$ for
$\sigma_{zz}^{max}=0.8\,\varepsilon\sigma^{-3}$ (orange), and by
$-0.10\,\varepsilon$ for
$\sigma_{zz}^{max}=1.0\,\varepsilon\sigma^{-3}$ (brown).  The data
for $\sigma_{zz}^{max}=0$ are indicated by the black curve. The
inset shows the average $U_{min}$ as a function of the stress
amplitude $\sigma_{zz}^{max}$.  The period of oscillations is
$T=500\,\tau$ and the temperature is $T_{LJ}=0.1\,\varepsilon/k_B$.}
\label{fig:poten_time_Tr1}
\end{figure}

The variation of the potential energy as a function of time is shown
in Fig.\,\ref{fig:poten_time_Tr1} for the stress amplitudes
$\sigma_{zz}^{max}\,\sigma^3/\varepsilon=0$, $0.2$, $0.4$, $0.6$,
$0.8$, and $1.0$.   It becomes apparent, when plotted on the log
scale, that in all cases the potential energy decreases
monotonically during the whole time interval of 5000 cycles, thus
slowly approaching the level of energy minimum. Note that similar
results for the potential energy series were reported for
periodically sheared glasses at constant volume and sufficiently
high temperatures $T_{LJ}\geqslant0.01\,\varepsilon/k_B$ in the
elastic range~\cite{Priezjev18,Priezjev18a}, while at lower
temperatures and in athermal systems, the potential energy levels
off after about 100 shear cycles near the critical strain
amplitude~\cite{Priezjev18,Sastry13,Reichhardt13}.   Typically, the
number of cycles required to reach a steady state decreases when the
strain amplitude is reduced away from the critical
value~\cite{Sastry13,Reichhardt13,Sastry17,Priezjev18}.

\vskip 0.05in


The upper black curve in Fig.\,\ref{fig:poten_time_Tr1} denotes the
data for $\sigma_{zz}^{max}=0$, \textit{i.e.}, it represents a
physical aging process at constant pressure applied in all three
dimensions, which leads to progressively lower energy states and
higher glass density (discussed below).   It was demonstrated
previously that, when the temperature of the KA mixture is reduced
below the glass transition and the system is allowed to evolve at
constant volume, the potential energy decreases as a power-law
function of time and the two-times correlation functions exhibit
strong time and waiting time dependence~\cite{KobBar97,KobBar00}.
Furthermore, as shown in Fig.\,\ref{fig:poten_time_Tr1}, with
increasing stress amplitude, the minimum of the potential energy
becomes deeper and the amplitude of the energy oscillations
increases.   Note that in order to avoid overlap between different
curves, the data are displaced vertically in
Fig.\,\ref{fig:poten_time_Tr1}.   The average value of the potential
energy after 5000 cycles was estimated as follows.   After the last
cycle, the pressure was set to zero in all three dimensions
(\textit{i.e.}, $\sigma_{zz}^{max}=0$) and the system was allowed to
relax during $10^4\tau$.  Thus, the average value of the potential
energy during the time interval $10^4\tau$ is reported in the inset
to Fig.\,\ref{fig:poten_time_Tr1}. It can be seen that periodic
loading with larger stress amplitudes results in lower values of the
potential energy.

\vskip 0.05in

\begin{figure}[t]
\includegraphics[width=12.cm,angle=0]{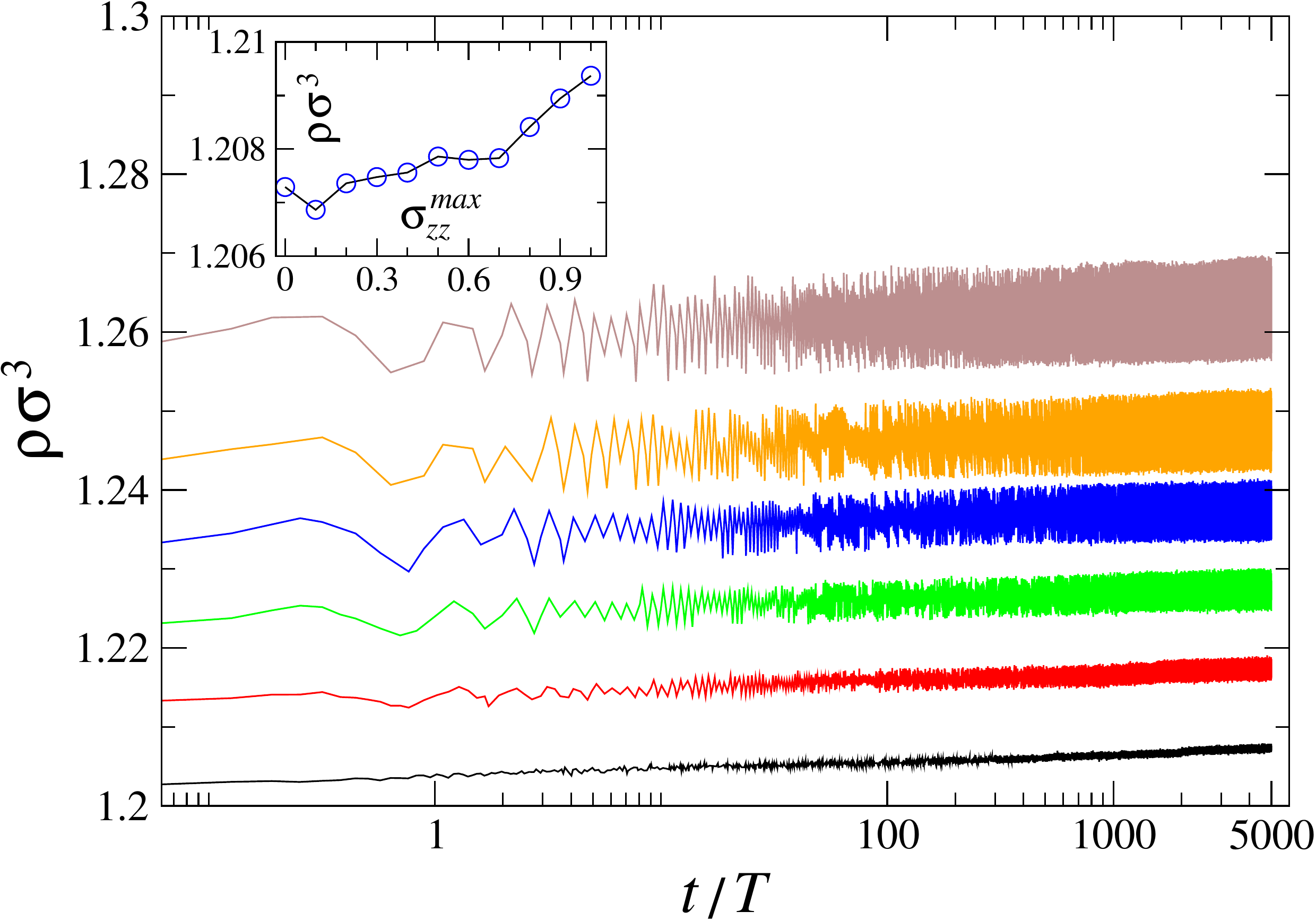}
\caption{(Color online) The time dependence of the glass density,
$\rho\sigma^3$, during 5000 cycles for the stress amplitudes
$\sigma_{zz}^{max}\,\sigma^3/\varepsilon=0.0$, $0.2$, $0.4$, $0.6$,
$0.8$, and $1.0$ (from bottom to top). For clarity, the data are
displaced vertically by $+0.01\,\sigma^3$ for
$\sigma_{zz}^{max}=0.2\,\varepsilon\sigma^{-3}$ (red), by
$+0.02\,\sigma^3$ for
$\sigma_{zz}^{max}=0.4\,\varepsilon\sigma^{-3}$ (green), by
$+0.03\,\sigma^3$ for
$\sigma_{zz}^{max}=0.6\,\varepsilon\sigma^{-3}$ (blue), by
$+0.04\,\sigma^3$ for
$\sigma_{zz}^{max}=0.8\,\varepsilon\sigma^{-3}$ (orange), and by
$+0.055\,\sigma^3$ for
$\sigma_{zz}^{max}=1.0\,\varepsilon\sigma^{-3}$ (brown). The inset
shows the average glass density after 5000 cycles as a function of
the stress amplitude (see text for details).}
\label{fig:density_time_Tr1}
\end{figure}

Next, the time evolution of the average glass density is plotted in
Fig.\,\ref{fig:density_time_Tr1} for the indicated stress
amplitudes. It is evident that both quiescent glass at constant
pressure and periodically stressed glasses become more dense over
time.   Similar to the potential energy, the average glass density
was extracted from the data during the additional time interval of
$10^4\tau$ at constant (zero) pressure after $5000$ cycles.   The
results are summarized in the inset in
Fig.\,\ref{fig:density_time_Tr1}. It can be observed that the
long-time average density becomes larger with increasing strain
amplitude.   Over many cycles, the more efficient packing at higher
strain amplitudes is achieved because groups of atoms can rearrange
more easily when a system is under tension, thus leading to lower
energy states at higher densities. Note, however, that the glass
density at zero pressure in the quiescent sample is slightly lower
than at the strain amplitude
$\sigma_{zz}^{max}\,\sigma^3/\varepsilon=0.01$, as shown in the
inset to Fig.\,\ref{fig:density_time_Tr1}.  Most probably this
behavior can be ascribed to a particular realization of disorder.

\vskip 0.05in

\begin{figure}[t]
\includegraphics[width=12.cm,angle=0]{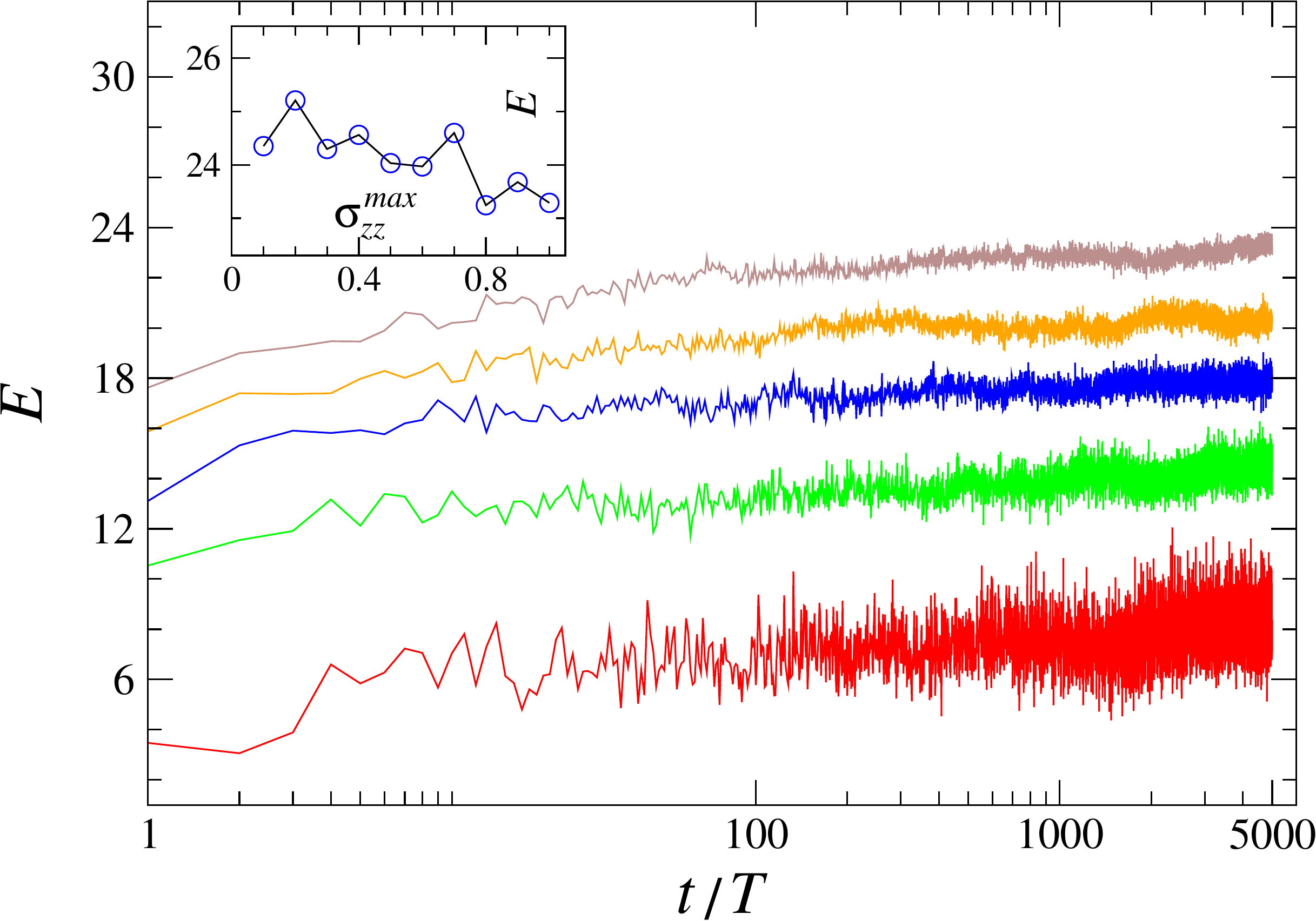}
\caption{(Color online) The variation of the elastic modulus, $E$
(in units of $\varepsilon\sigma^{-3}$), during 5000 cycles for the
stress amplitudes $\sigma_{zz}^{max}\,\sigma^3/\varepsilon=0.2$,
$0.4$, $0.6$, $0.8$, and $1.0$ (from bottom to top).    The data are
shifted by $-17.0\,\varepsilon\sigma^{-3}$ for
$\sigma_{zz}^{max}\,\sigma^3/\varepsilon=0.2$ (red), by
$-10.0\,\varepsilon\sigma^{-3}$ for
$\sigma_{zz}^{max}\,\sigma^3/\varepsilon=0.4$ (green), by
$-6.0\,\varepsilon\sigma^{-3}$ for
$\sigma_{zz}^{max}\,\sigma^3/\varepsilon=0.6$ (blue), and by
$-3.0\,\varepsilon\sigma^{-3}$ for
$\sigma_{zz}^{max}\,\sigma^3/\varepsilon=0.8$ (orange). The brown
curve indicates the original data for
$\sigma_{zz}^{max}\,\sigma^3/\varepsilon=1.0$.  The inset shows $E$
after 5000 cycles versus $\sigma_{zz}^{max}$.   }
\label{fig:modulus_time_Tr1}
\end{figure}

The time dependence of the elastic modulus, $E$, is reported in
Fig.\,\ref{fig:modulus_time_Tr1} for the selected strain amplitudes.
In our study, the modulus was computed simply as a ratio of the
stress amplitude and the strain amplitude for every cycle. As can be
seen, $E$ increases on average as a function of time for each stress
amplitude.  Note that the data are more noisy at smaller stress
amplitudes.  Furthermore, the inset in
Fig.\,\ref{fig:modulus_time_Tr1} shows $E$, averaged over the last
100 cycles, versus the stress amplitude $\sigma_{zz}^{max}$.
Although the data are somewhat scattered, the trend is evident;
namely, the elastic modulus decreases at larger stress amplitudes.
These results are consistent with conclusions from the previous MD
study on periodically sheared binary glasses at constant volume,
where it was found that, with increasing strain amplitude up to a
critical value, the potential energy decreases but at the same time
the storage modulus is also reduced~\cite{Priezjev18a}.   This trend
can be rationalized by realizing that, when the strain amplitude
increases, nonaffine displacements of atoms become increasingly
broadly distributed in strained samples, leading to a local stress
relaxation~\cite{Priezjev16a}.

\vskip 0.05in


We next analyze the spatial distribution of atoms with large
nonaffine displacements after a full cycle. To remind, the nonaffine
measure is computed numerically using the transformation matrix
$\mathbf{J}_i$, which linearly transforms the positions of
neighboring atoms during the time interval $\Delta t$ and minimizes
the quantity $D^2(t, \Delta t)$ in the following
expression~\cite{Falk98}:
\begin{equation}
D^2(t, \Delta t)=\frac{1}{N_i}\sum_{j=1}^{N_i}\Big\{
\mathbf{r}_{j}(t+\Delta t)-\mathbf{r}_{i}(t+\Delta t)-\mathbf{J}_i
\big[ \mathbf{r}_{j}(t) - \mathbf{r}_{i}(t)    \big] \Big\}^2,
\label{Eq:D2min}
\end{equation}
where the sum is taken over nearest neighbors within the cutoff
distance of $1.5\,\sigma$ from the position of the $i$-th atom
$\mathbf{r}_{i}(t)$. Based on the previous analysis, it is expected
that at sufficiently low temperatures, the system starts to deform
reversibly after a number of transient cycles, and the nonaffine
displacements of atoms after a full cycle will be smaller than the
cage size~\cite{Priezjev18}. In contrast, above the critical strain
amplitude, it was found that in steady state of deformation at
finite temperature, a significant fraction of atoms undergo large
nonaffine displacements after a full cycle and form a
system-spanning shear band in both well~\cite{Priezjev17} and
poorly~\cite{Priezjev18a} annealed glasses.

\vskip 0.05in

\begin{figure}[t]
\includegraphics[width=12.cm,angle=0]{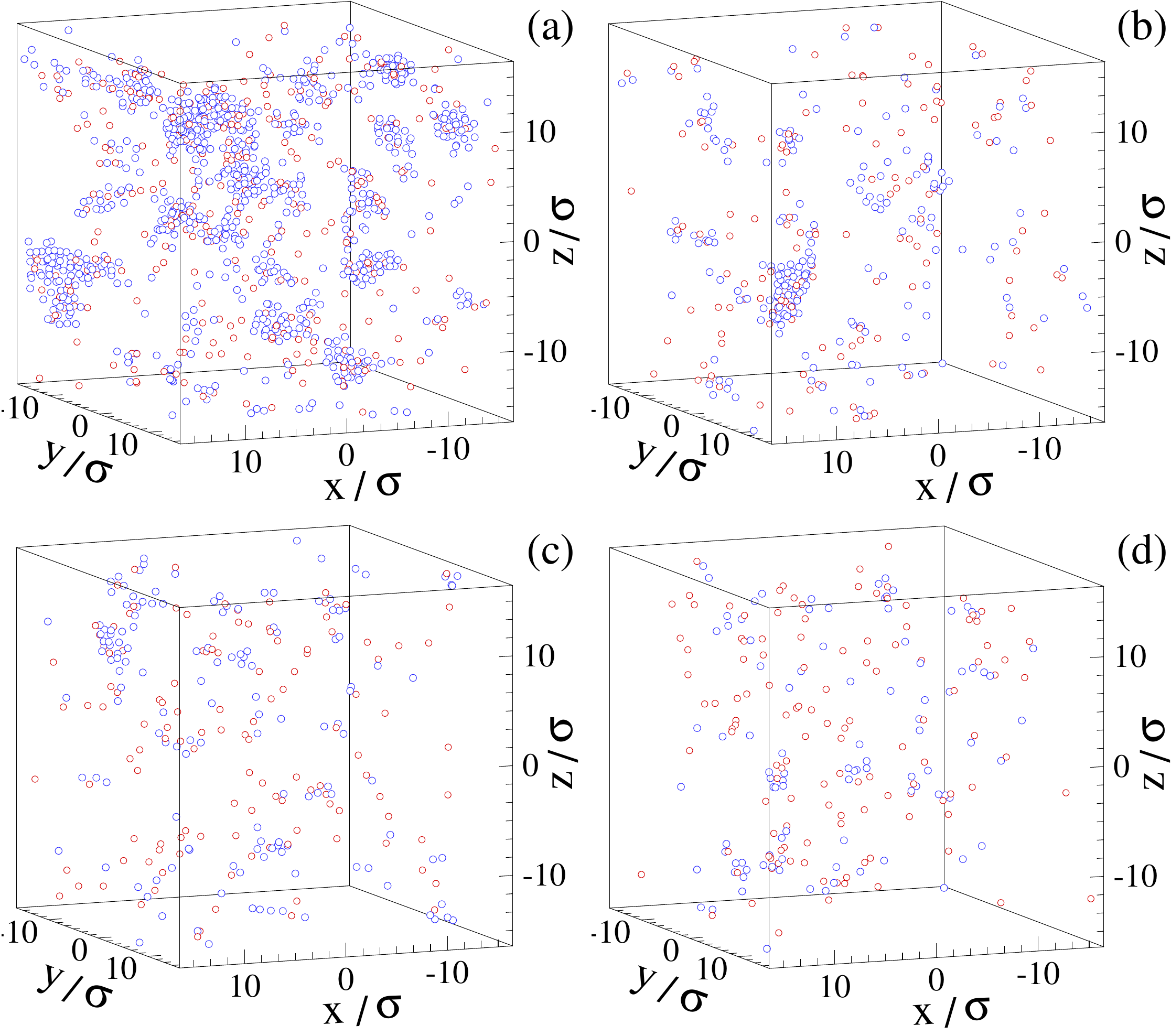}
\caption{(Color online) Atomic configurations for the stress
amplitude $\sigma_{zz}^{max}=0.2\,\varepsilon\sigma^{-3}$ and
nonaffine measure (a) $D^2(9T,T)>0.04\,\sigma^2$, (b)
$D^2(99T,T)>0.04\,\sigma^2$, (c) $D^2(999T,T)>0.04\,\sigma^2$, and
(d) $D^2(4999T,T)>0.04\,\sigma^2$.}
\label{fig:snapshot_clusters_dp02_Tr1}
\end{figure}
\begin{figure}[t]
\includegraphics[width=12.cm,angle=0]{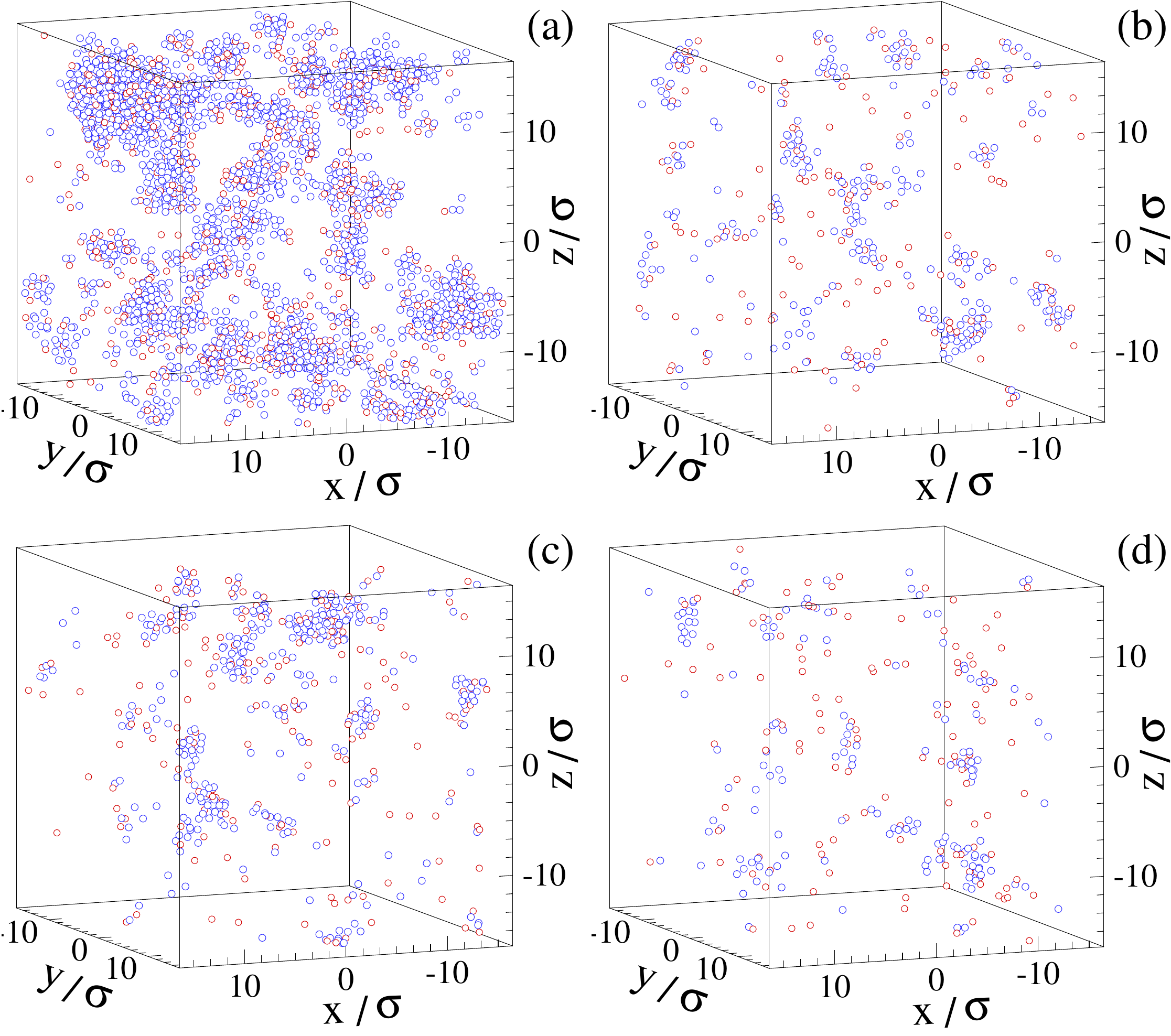}
\caption{(Color online) Snapshots of atomic positions for the stress
amplitude $\sigma_{zz}^{max}=1.0\,\varepsilon\sigma^{-3}$ and
nonaffine measure (a) $D^2(9T,T)>0.04\,\sigma^2$, (b)
$D^2(99T,T)>0.04\,\sigma^2$, (c) $D^2(999T,T)>0.04\,\sigma^2$, and
(d) $D^2(4999T,T)>0.04\,\sigma^2$.}
\label{fig:snapshot_clusters_dp10_Tr1}
\end{figure}

The consecutive snapshots of atomic positions with large nonaffine
displacements are displayed in
Fig.\,\ref{fig:snapshot_clusters_dp02_Tr1} for the stress amplitude
$\sigma_{zz}^{max}=0.2\,\varepsilon\sigma^{-3}$ and in
Fig.\,\ref{fig:snapshot_clusters_dp10_Tr1} for
$\sigma_{zz}^{max}=1.0\,\varepsilon\sigma^{-3}$.   In each case, the
nonaffine measure, Eq.\,(\ref{Eq:D2min}), was computed for atomic
configurations at the beginning and at the end of a cycle when the
applied stress is zero, \textit{i.e.}, $\Delta t=T$.   It can be
seen from Figs.\,\ref{fig:snapshot_clusters_dp02_Tr1}\,(a) and
\ref{fig:snapshot_clusters_dp10_Tr1}\,(a) that after the first ten
oscillation periods following the thermal quench, the atoms with
relatively large nonaffine displacements,
$D^2(9T,T)>0.04\,\sigma^2$, form a number of compact clusters. After
hundreds of cycles, the irreversible rearrangements become rare and
they mostly consist of isolated smaller (type $B$) atoms that are
more mobile.  We note that the presence of a finite number of atoms
with large nonaffine displacements after 5000 cycles [\,see
Figs.\,\ref{fig:snapshot_clusters_dp02_Tr1}\,(d) and
\ref{fig:snapshot_clusters_dp10_Tr1}\,(d)\,] is consistent with the
continuing decay of the potential energy and glass densification
shown in Figs.\,\ref{fig:poten_time_Tr1} and
\ref{fig:density_time_Tr1}, respectively.   In other words, as the
number of cycles increases up to $5000$, the system continues to
explore states with lower potential energy via occasional
irreversible rearrangements of small groups of atoms.   Finally, the
absence of a shear band that consists of atoms with large nonaffine
displacements during the time interval $5000\,T$ confirms that the
system deforms elastically even at the largest stress amplitude
considered in the present study.

\section{Conclusions}

In summary, the influence of the stress-controlled,
tension-compression cyclic loading on mechanical properties of
binary glasses was examined using molecular dynamics simulations.
The model glass was represented by a non-additive binary mixture,
which was rapidly cooled from a liquid state to a temperature well
below the glass transition. The uniaxial tension-compression loading
was applied periodically, while the pressure in lateral directions
was kept constant. In such a setup, during thousands of cycles in
the elastic regime, the systems continue relocating into the states
with deeper potential energy minima and gradually become denser and
stronger.  It was found that under cycling loading with larger
stress amplitudes, the samples become more dense, the minimum of the
potential energy is lower, and the elastic modulus is reduced.
Finally, the slow relaxation process is accompanied with the
appearance of clusters of atoms with large nonaffine displacements,
whose typical size becomes smaller as the number of cycles
increases.

\section*{Acknowledgments}

Financial support from the National Science Foundation (CNS-1531923)
is gratefully acknowledged. The article was prepared within the
framework of the Basic Research Program at the National Research
University Higher School of Economics (HSE) and supported within the
framework of a subsidy by the Russian Academic Excellence Project
`5-100'. The molecular dynamics simulations were performed using the
LAMMPS numerical code developed at Sandia National
Laboratories~\cite{Lammps}. Computational work in support of this
research was performed at Wright State University's Computing
Facility and the Ohio Supercomputer Center.


%
\begin{figure}[t]
\includegraphics[width=10.0cm,angle=0]{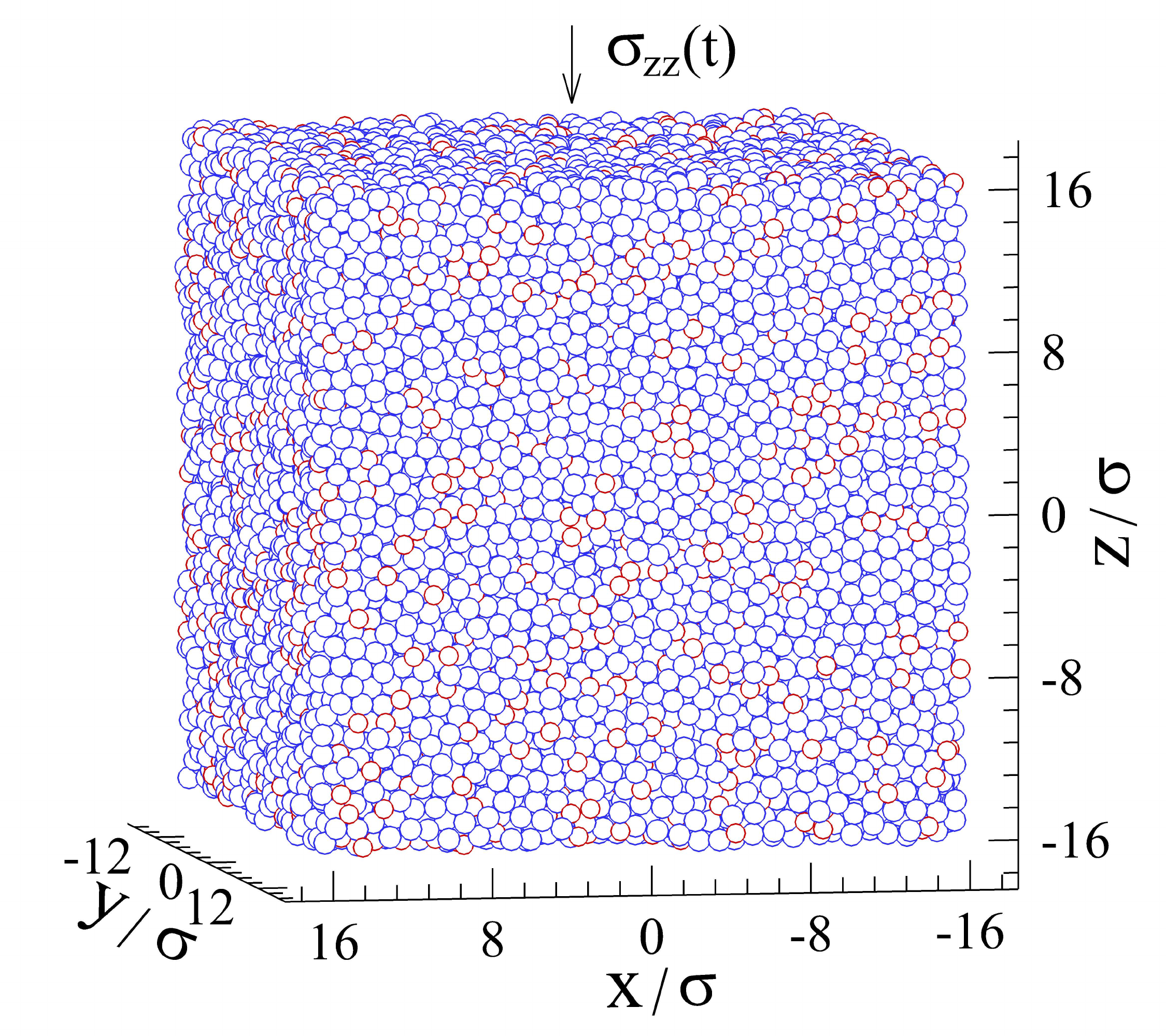}
\caption{(Color online) A snapshot of the Lennard-Jones binary glass
($N_{tot}=40\,000$) after the quench to the temperature
$T_{LJ}=0.1\,\varepsilon/k_B$. The time-dependent stress
$\sigma_{zz}(t)$ is applied along the $\hat{z}$ direction, while the
normal stresses along the $\hat{x}$ and $\hat{y}$ directions are
kept constant (see text for details). Note that atoms of types $A$
and $B$ (indicated by blue and red circles) are now shown to scale.}
\label{fig:snapshot_system}
\end{figure}

%

%

%

%

%

%

\bibliographystyle{prsty}

\begin{thebibliography}{99}


\bibitem{WangRev12}    W.~H. Wang,
                       The elastic properties, elastic models and elastic perspectives of metallic glasses,
                       Prog. Mater. Sci. {\bf 57}, 487 (2012).

\bibitem{Argon79}      A.~S. Argon,
                       Plastic deformation in metallic glasses,
                       Acta Metall. {\bf 27}, 47 (1979).

\bibitem{Spaepen77}    F. Spaepen,
                       A microscopic mechanism for steady state inhomogeneous flow in metallic glasses,
                       Acta Metall. {\bf 25}, 407 (1977).

\bibitem{Falk98}       M.~L. Falk and J.~S. Langer,
                       Dynamics of viscoplastic deformation in amorphous solids,
                       Phys. Rev. E {\bf 57}, 7192 (1998).

\bibitem{EvanMa16}     J. Ding, Y.-Q. Cheng, H. Sheng, M. Asta, R.~O. Ritchie, and E. Ma,
                       Universal structural parameter to quantitatively predict metallic glass properties,
                       Nat. Commun. {\bf 7}, 13733 (2016).



\bibitem{Schuh08}      C.~E. Packard, L.~M. Witmer, and C.~A. Schuh,
                       Hardening of a metallic glass during cyclic loading in the elastic range,
                       Appl. Phys. Lett. {\bf 92}, 171911 (2008).

\bibitem{Schuh12}      C. Deng and C.~A. Schuh,
                       Atomistic mechanisms of cyclic hardening in metallic glass,
                       Appl. Phys. Lett. {\bf 100}, 251909 (2012).


\bibitem{WangASS17}    D. Zhao, H. Zhao, B. Zhu, and S. Wang,
                       Investigation on hardening behavior of metallic glass under cyclic
                       indentation loading via molecular dynamics simulation,
                       Appl. Surf. Sci. {\bf 416}, 14 (2017).

\bibitem{Lashgari18}   H.~R. Lashgari, C. Tang, D. Chu, and S. Li,
                       Molecular dynamics simulation of cyclic indentation in Fe-based amorphous alloy,
                       Comput. Mater. Sci. {\bf 143}, 473 (2018).



\bibitem{GaoNano15}    Z.~D. Sha, S.~X. Qu, Z.~S. Liu, T.~J. Wang, and H. Gao,
                       Cyclic deformation in metallic glasses,
                       Nano Lett. {\bf 15}, 7010 (2015).

\bibitem{GaoSize17}    Z. Sha, W.~H. Wong, Q. Pei, P.~S. Branicio, Z. Liu, T. Wang, T. Guo, H. Gao,
                       Atomistic origin of size effects in fatigue behavior of metallic glasses,
                       J. Mech. Phys. Solids {\bf 104}, 84 (2017).

\bibitem{Liaw10}       Y.~C. Lo, H.~S. Chou, Y.~T. Cheng, J.~C. Huang, J.~R. Morris, and P.~K. Liaw,
                       Structural relaxation and self-repair behavior in nano-scaled Zr–Cu metallic glass
                       under cyclic loading: Molecular dynamics simulations,
                       Intermetallics {\bf 18}, 954 (2010).

\bibitem{Shi11}        Y. Shi,  D. Louca, G. Wang, and P.~K. Liaw,
                       Compression-compression fatigue study on model metallic glass
                       nanowires by molecular dynamics simulations,
                       J. Appl. Phys. {\bf 110}, 023523 (2011).


\bibitem{Shi15}        J. Luo, K. Dahmen, P.~K. Liaw, and Y. Shi,
                       Low-cycle fatigue of metallic glass nanowires,
                       Acta Mater. {\bf 87}, 225 (2015).


\bibitem{Yang16}       Y.~F. Ye, S. Wang, J. Fan, C.~T. Liu, and Y. Yang,
                       Atomistic mechanism of elastic softening in metallic glass under cyclic
                       loading revealed by molecular dynamics simulations,
                       Intermetallics {\bf 68}, 5 (2016).




\bibitem{Sastry13}     D. Fiocco, G. Foffi, and S. Sastry,
                       Oscillatory athermal quasistatic deformation of a model glass,
                       Phys. Rev. E {\bf 88}, 020301(R) (2013).

\bibitem{Reichhardt13} I. Regev, T. Lookman, and C. Reichhardt,
                       Onset of irreversibility and chaos in amorphous solids under periodic shear,
                       Phys. Rev. E {\bf 88}, 062401 (2013).

\bibitem{IdoNature15}  I. Regev, J. Weber, C. Reichhardt, K.~A. Dahmen, and T. Lookman,
                       Reversibility and criticality in amorphous solids,
                       Nat. Commun. {\bf 6}, 8805 (2015).

\bibitem{Kawasaki16}   T. Kawasaki and L. Berthier,
                       Macroscopic yielding in jammed solids is accompanied by a non-equilibrium
                       first-order transition in particle trajectories,
                       Phys. Rev. E {\bf 94}, 022615 (2016).

\bibitem{Sastry17}     P. Leishangthem, A.~D.~S. Parmar, and S. Sastry,
                       The yielding transition in amorphous solids under oscillatory shear deformation,
                       Nat. Commun. {\bf 8}, 14653 (2017).

\bibitem{Lavrentovich17} M.~O. Lavrentovich, A.~J. Liu, and S.~R. Nagel,
                         Period proliferation in periodic states in cyclically sheared jammed solids,
                         Phys. Rev. E {\bf 96}, 020101(R) (2017).

\bibitem{OHern17}      M. Fan, M. Wang, K. Zhang, Y. Liu, J. Schroers, M.~D. Shattuck, and C.~S. O'Hern,
                       The effects of cooling rate on particle rearrangement statistics:
                       Rapidly cooled glasses are more ductile and less reversible,
                       Phys. Rev. E {\bf 95}, 022611 (2017).


\bibitem{Priezjev13}   N.~V. Priezjev,
                       Heterogeneous relaxation dynamics in amorphous materials under cyclic loading,
                       Phys. Rev. E {\bf 87}, 052302 (2013).

\bibitem{Priezjev14}   N.~V. Priezjev,
                       Dynamical heterogeneity in periodically deformed polymer glasses,
                       Phys. Rev. E {\bf 89}, 012601 (2014).

\bibitem{Priezjev16}   N.~V. Priezjev,
                       Reversible plastic events during oscillatory deformation of amorphous solids,
                       Phys. Rev. E {\bf 93}, 013001 (2016).

\bibitem{Priezjev16a}  N.~V. Priezjev,
                       Nonaffine rearrangements of atoms in deformed and quiescent binary glasses,
                       Phys. Rev. E {\bf 94}, 023004 (2016).


\bibitem{Priezjev17}   N.~V. Priezjev,
                       Collective nonaffine displacements in amorphous materials during large-amplitude oscillatory shear,
                       Phys. Rev. E {\bf 95}, 023002 (2017).

\bibitem{Hecke17}      S. Dagois-Bohy, E. Somfai, B.~P. Tighe, and M. van Hecke,
                       Softening and yielding of soft glassy materials,
                       Soft Matter {\bf 13}, 9036 (2017).

\bibitem{Keblinsk17}   R. Ranganathan, Y. Shi, and P. Keblinski,
                       Commonalities in frequency-dependent viscoelastic damping in glasses
                       in the MHz to THz regime,
                       J. Appl. Phys. {\bf 122}, 145103 (2017).

\bibitem{Priezjev18}   N.~V. Priezjev,
                       Molecular dynamics simulations of the mechanical annealing process in
                       metallic glasses: Effects of strain amplitude and temperature,
                       J. Non-Cryst. Solids {\bf 479}, 42 (2018).

\bibitem{Priezjev18a}  N.~V. Priezjev,
                       The yielding transition in periodically sheared binary glasses at finite temperature,
                       Comput. Mater. Sci. {\bf 150}, 162 (2018).




\bibitem{KobAnd95}     W. Kob and H.~C. Andersen,
                       Testing mode-coupling theory for a supercooled binary Lennard-Jones mixture:
                       The van Hove correlation function,
                       Phys. Rev. E {\bf 51}, 4626 (1995).

\bibitem{Weber85}      T.~A. Weber and F.~H. Stillinger,
                       Local order and structural transitions in amorphous metal-metalloid alloys,
                       Phys. Rev. B {\bf 31}, 1954 (1985).


\bibitem{Allen87}      M.~P. Allen and D.~J. Tildesley,
                       {\it Computer Simulation of Liquids} (Clarendon, Oxford, 1987).


\bibitem{Lammps}       S.~J. Plimpton,
                       Fast parallel algorithms for short-range molecular dynamics,
                       J. Comp. Phys. {\bf 117}, 1 (1995).

\bibitem{KobBar97}     W. Kob and J.-L. Barrat,
                       Aging Effects in a Lennard-Jones Glass,
                       Phys. Rev. Lett. {\bf 78}, 4581 (1997).

\bibitem{KobBar00}     W. Kob and J.-L. Barrat,
                       Fluctuations, response and aging dynamics in a simple
                       glass-forming liquid out of equilibrium,
                       Eur. Phys. J. B {\bf 13}, 319 (2000).

\end{thebibliography}

\end{document}